# Low RCS High-Gain Broadband Substrate Integrated Waveguide Antenna Based on Elliptical Polarization Conversion Metasurface

Cuiqin Zhao, Dongya Shen, *Member, IEEE* , Yanming Duan, Yuting Wang, Huihui Xiao, Longxiang Luo

*Abstract*—Designed an elliptical polarization conversion metasurface (PCM) for Ka-band applications, alongside a high-gain substrate integrated waveguide (SIW) antenna. The PCM elements are integrated into the antenna design in a chessboard array configuration, with the goal of achieving effective reduction in the antenna's radar cross section (RCS). Both the PCM elements and antenna structure exhibit a simple design. The top layer of the metasurface (MS) elements employs an elliptical pattern symmetric along the diagonal, enabling efficient conversion of linearly polarized waves. The antenna component, on the other hand, consists of a broadband dipole antenna fed by SIW slot coupling. Verified through simulations, the polarization conversion bandwidth of this PCM unit reaches 80.38% where polarization conversion ratio (PCR) exceeds 90% (25.3-59.3GHz), demonstrating exceptional conversion performance. When the dipole antenna is combined with the PCM, its -10dB impedance bandwidth reaches to 15.09% (33.7-39.2GHz), with a maximum realized gain of 9.1dBi. Notably, the antenna loaded with the chessboard PCM structure effectively disperses the energy of scattered echoes around, significantly reducing the concentration of scattered energy in the direction of the incident wave, thereby achieving an effective reduction in RCS.

## I. INTRODUCTION

WITH the rapid advancement of wireless communication technology, the Ka-band, with its high-frequency characteristics, has demonstrated crucial application value in cutting-edge fields such as satellite communication and radar detection. However, high-frequency signals are highly susceptible to polarization mismatch and radar cross section (RCS) during transmission, challenges that directly limit the overall efficiency of communication systems. To mitigate communication blackout area, antennas are typically positioned in conspicuous locations, resulting in their inherent strong scattering characteristics [1]. Consequently, exploring effective methods to reduce the RCS of antennas has emerged as a critical and urgent research topic in stealth technology [2][3][4]. In response to this issue, various strategies have been developed to lower antenna RCS, including but not limited to designing antenna configurations with low scattering properties[3][4], incorporating reconfigurable technology for flexible antenna performance adjustment[5][6][7][8][9][10][11][12], leveraging Pancharatnam–Berry (PB) phase gradients[13] and absorbing materials [14][15][16], and employing phase cancellation techniques through polarization conversion metasurfaces (PCMs)[17][18][19][20][21][22].

The advantage of designing an antenna with RCS reduction characteristics [3][4] lies in achieving RCS reduction without the need for additional structures. However, this design typically requires the construction of two different structures with similar amplitudes but opposite phases to realize RCS reduction. Unfortunately, such designs often suffer from narrow RCS reduction bandwidth, and the reduction effect is limited to specific in-band or out-of-band frequencies. Furthermore, the coupling between these two different structures can adversely affect the antenna's radiation performance. In [3], the introduction of via-hole design was used to broaden the reduction bandwidth and reduce coupling, but this also increased the structural complexity. On the other hand, Frequency Selective Surfaces (FSS) and absorbing materials [5][6][7][10][11][12][14][16] can also achieve RCS reduction. Nevertheless, these designs often rely on components such as p-i-n diodes to construct different structures and directly achieve a 180° phase difference. Such designs increase the structural complexity of the antenna system, potentially leading to higher manufacturing costs and increased system integration difficulty. It is worth noting that the phase difference achieved in [11] was 180°±37°, which deviates from the ideal 180°, potentially affecting polarization purity. Similarly, arranging two opposite phases artificial magnetic conductors (AMCs) in a checkerboard pattern [10] can effectively reduce RCS. However, the coupling effect between the electromagnetic band gap (EBG) structure and the

Manuscript received December 29, 2024. This work was supported in part by the National Natural Science Foundation of China (62361057), the Guangxi Provincial Department of Education(2023KY0630, 2023KY0636) and the 2024 Undergraduate Innovation and Entrepreneurship Projects (202410605024, S202410605026). (Corresponding author: Dongya Shen.)

Cuiqin Zhao, Yanming Duan, Yuting Wang, Huihui Xiao, Longxiang Luo are with College of Big Data and Computer Science, Hechi University, Yizhou 546300, China (e-mail: 07027@hcnu.edu.cn, 364214497@qq.com, 767744526@qq.com, 317219459@qq.com, 77428798@qq.com).

Dongya Shen, Cuiqin Zhao are with the Yunnan Provincial Engineering Laboratory of Cloud Wireless Access & Heterogeneous Networks, Yunnan University, Kunming 650091, China (e-mail: shendy@ynu.edu.cn).

antenna patch can interfere with the antenna's radiation performance. To address this issue, [10] proposed an improved solution by setting a certain height gap between the antenna and the AMC. Although this helps to isolate the two, it also correspondingly increases the antenna's profile height.

As an emerging artificial electromagnetic (EM) material, PCM exhibits immense potential in the field of RCS reduction due to its exceptional EM manipulation capabilities [18][19][20][21][22][27]. By intelligently manipulating the polarization direction of EM waves, PCM achieves efficient scattering and absorption, significantly lowering the target's RCS and substantially enhancing the stealth performance of EM devices. Furthermore, by integrating PCM units with their mirror elements in a checkerboard layout, a successful wideband RCS reduction has been achieved, paving a new path for the development of stealth technology.

Although numerous antennas with reduced RCS have been proposed, achieving antennas that are simultaneously high-frequency, miniaturized, and bandwidth-enhanced remains a challenge. In this paper, a low-RCS, broadband, and miniaturized SIW antenna with elliptical PCM is introduced. The main contributions are as follows:

1) Previously, achieving a 180° phase difference between adjacent structures is difficult when using multiple different structures for scattering, often resulting in errors of about 30°. In this work, a checkerboard structured PCM array is employed, which easily ensures that the phase difference between the PCM unit and its mirrored unit is almost exactly 180°. This ensures high polarization purity and a low RCS reduction value.

2) Compared to previous PCMs, this paper utilizes a simple elliptical structure with few parameters to achieve polarization conversion. By merely adjusting the axial ratio of the ellipse, the polarization conversion characteristics can be effectively controlled. The polarization conversion bandwidth, with a polarization conversion ratio (PCR) exceeding 90%, reaches 80.38% (25.3 - 59.3 GHz). The phase difference between the PCM unit and its mirrored unit remains at 180° across a frequency range from 20 GHz to 60 GHz, thereby enabling RCS reduction both in-band and out of band.

3) The radiating part of the antenna adopts an EM dipole design fed through SIW slot coupling, with overall dimensions of only 11.6mm x 8mm x 1.762mm. The antenna achieves a -10dB impedance bandwidth of 24.4% (28.8-36.8 GHz) and a maximum realized gain of up to 8.4 dBi. Therefore, this antenna combines the characteristics of miniaturization, high gain, and broadband performance.

The remainder of this paper is structured as follows: Section II provides a detailed design and mechanism analysis, including the structure of the PCM unit, the principle behind RCS reduction using PCM units, the configuration of the SIW antenna, and the integration of the antenna with PCMs. Section III focuses on the antenna performances, specifically examining the radiation and scattering performance through both simulation and experimental results. Lastly, Section IV presents the concluding remarks.

## II. DESIGN AND MECHANISM ANALYSIS

### A. Structure of the PCM unit

The PCM unit and its geometric dimensions are illustrated in Fig. 1. Fig. 1(a) depicts the top view, Fig. 1(b) shows the side view, and Fig. 1(c) displays the checkerboard array formed by the PCM unit and its mirror unit. The PCM unit is structured in three layers, with the middle layer featuring a lossy Rogers RT-duroid 5880 substrate with a permittivity of 2.2 and a loss tangent of 0.0009. The top layer consists of metallic ellipses that are symmetrical along the diagonal, while the bottom layer is a continuous copper-clad ground plane. To further investigate these properties, the advanced EM simulation tool CST is employed to conduct a detailed simulation analysis of the designed reflective PCM unit. The optimized parameters (in mm) are as follows: tm=0.035, Ra=3.8, Rb=1.3, Pms=4, H1=1.

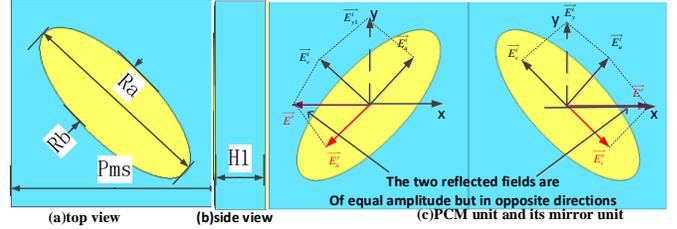

**Fig. 1.** Structure of PCM

### B. Principle of RCS reduction by PCM units

To gain a deeper understanding of the polarization conversion and phase cancellation characteristics of the chessboard structured PCM, consider the following scenario: A plane EM wave is normally incident on the PCM, with its electric field vector assumed to be polarized along the y-axis. Consequently, the incident EM wave can be represented as $\overrightarrow{E_y^i}$. Further, $\overrightarrow{E_y^i}$ can be decomposed into two components: $\overrightarrow{E_u^i}$ along the long axis and $\overrightarrow{E_v^i}$ along the short axis, as illustrated in Fig. 1(c).When the EM wave is incident on the MS, the presence of elliptical conductors induces a specific EM response. By adjusting the axis ratio of the ellipse, the PCM unit can be made to exhibit significant electrical resonance along the long axis, where the long conductor is located, while the electrical resonance along the short axis, where the short conductor is, is relatively weak. Specifically, the strong resonance along the long axis results in the generation of a reflected field $\overrightarrow{E_u^r}$ that is opposite in direction to the incident field $\overrightarrow{E_u^i}$, implying a 180° phase difference between $\overrightarrow{E_u^i}$ and $\overrightarrow{E_u^r}$. In other words, the incident field $\overrightarrow{E_u^i}$ is largely reflected as $\overrightarrow{E_u^r}$. By contrast, due to the weak resonance along the short axis, its reflected field is negligible, and thus the field along the short axis remains primarily as the incident field component $\overrightarrow{E_v^i}$.

By vectorially synthesizing $\overrightarrow{E_u^r}$ and $\overrightarrow{E_v^i}$, the total reflected field $\overrightarrow{E^r}$ is obtained. At this point, there is a 90° phase difference between the incident field $\overrightarrow{E_y^i}$ and the reflected field $\overrightarrow{E^r}$, and their amplitudes are similar. This characteristic confirms that for an EM wave incident from the y-direction, its reflected field is reflected along the negative x-axis, thereby achieving cross-polarization conversion.

For the mirrored unit of the PCM, the situation is similar, but the polarization direction of the reflected field is opposite. That is, when a y-polarized wave is incident, the reflected field is along the positive x-axis. When the PCM unit and its mirrored unit are arranged in a chessboard pattern, they produce two reflected fields with equal amplitudes but opposite phases. Since these two reflected fields are located in the same plane and point in opposite directions, they mutually cancel each other out, achieving phase cancellation.

The situation for the incident electric field in the x-direction is similar to that in the y-direction and will not be repeated here.

To further validate the PCM unit's effectiveness in reducing RCS, Fig. 2 simulates reflective properties of the PCM unit. The reflection coefficients for both co-polarization and cross-polarization of the PCM unit are displayed in Fig. 2(a). To assess the efficiency of polarization conversion, the concept of the PCR is employed [23]. Using a y-polarized incident wave as a reference, the PCR is determined by the formula $PCR = r_{xy}^2 / \left( r_{xy}^2 + r_{yy}^2 \right)$ , where $r_{yy}$ and $r_{xy}$ denote the co-polarized and cross-polarized reflection coefficients, respectively. The PCR reveals that within the frequency band where PCR exceeds 90%, the PCM unit exhibits a polarization conversion bandwidth of 80.38% (25.3-59.3GHz), fully demonstrating its superior polarization conversion capability. Notably, within this operating frequency band, the cross-polarization reflection magnitude remains stably above 0.9, highlighting its strong cross-polarization reflection intensity. In contrast, the co-polarization reflection magnitude is relatively weak, further emphasizing the efficiency of polarization conversion. Four resonant frequencies of co-polarization reflection coefficients at 27.4, 43, 55 and 58.9GHz are achieved along with a wideband below −10 dB. Due to the symmetrical nature of the design, analogous results are observed for an x-polarized incident wave.

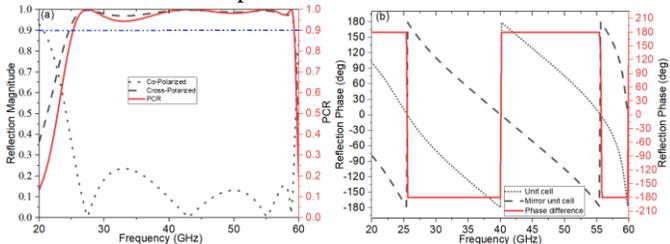

**Fig. 2.** Amplitude, PCR, phase and phase difference of the PCM

Fig. 2(b) simulates the phase variations of the PCM unit and its mirror unit at different frequencies, as well as the phase differences between them. Upon close observation, it can be seen that within the wide frequency range of 20-60GHz, the phase response of these two PCM units exhibits linear characteristics, consistently maintaining an almost perfect ±180° phase difference. This phenomenon strongly demonstrates the extremely high stability and consistency of the PCM units during the polarization conversion process. This experimental result strongly verifies the fundamental principles and effectiveness of the checkerboard PCM array in achieving RCS reduction.

Fig. 3 illustrates the surface current distributions on the top and bottom of the PCM unit at the four frequencies of 27.4, 35, 43 and 55 GHz. Upon examining Fig. 3(a) and (c), it is evident that the surface current directions on the top and bottom of the PCM unit are opposite, indicating the occurrence of magnetic resonance at these two frequency points. Fig. 3(b), on the other hand, shows that the current directions on the top and bottom are orthogonal, suggesting the occurrence of polarization conversion. Additionally, it can be observed from Fig. 3(d) that the current directions on the top and bottom are in-phase, resulting in electric resonance. Consequently, the designed PCM unit achieves broadband and efficient polarization conversion by integrating multiple mechanisms of electric resonance, magnetic resonance, and polarization conversion.

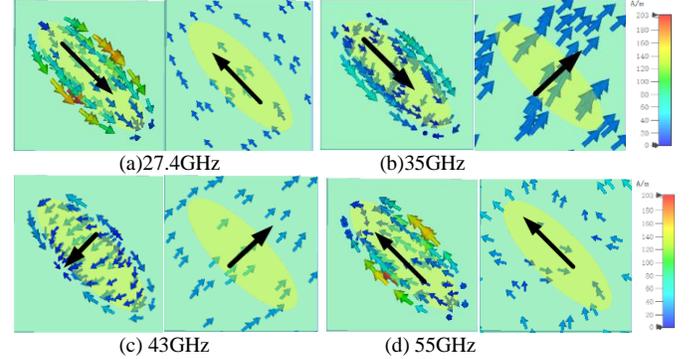

(a)27.4GHz  (b)35GHz

(c) 43GHz  (d) 55GHz

**Fig. 3.** Surface current distributions on the top and bottom of the PCM unit.

### C. Configuration of the SIW antenna

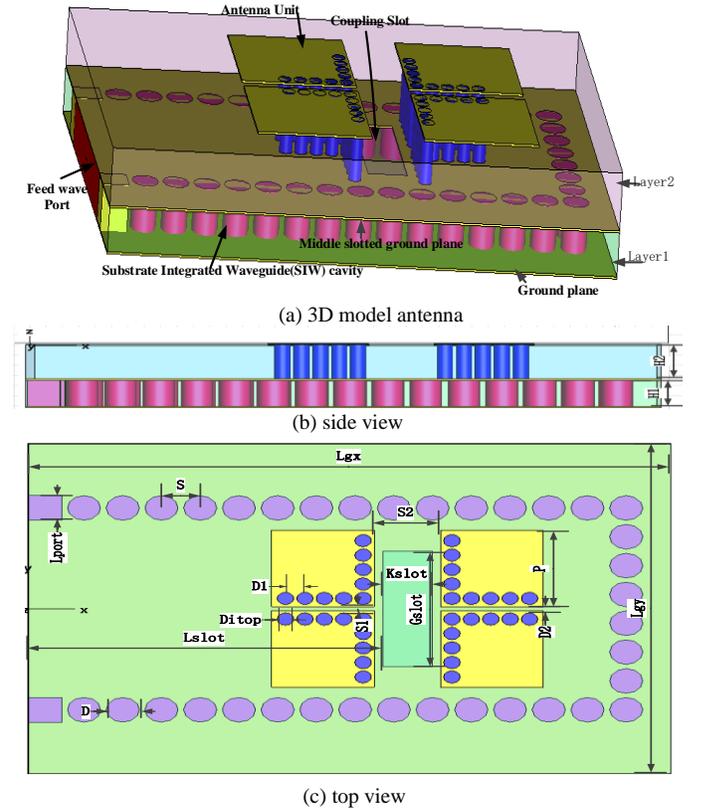

(a) 3D model antenna

(b) side view

(c) top view

**Fig. 4.** Schematic diagram of antenna structure

The structure of the SIW antenna is shown in Fig. 4, where Fig. 4(a) is 3D model, Fig. 4(b) is a side view and Fig. 4(c) is a top view. the antenna is comprised of two layers, possessing thicknesses of H1 and H2, respectively. Both the upper and lower dielectric layers of the antenna adopt the same dielectric substrate as the PCM. At the top, four radiation patches act as electric dipoles, while the blue copper holes function as magnetic dipoles. Between the two dielectric layers lies a slotted ground plane. The lower layer's red copper holes form a substrate integrated waveguide (SIW) cavity, which achieves coupled feeding through the middle slot.

To delve into the specific impact of antenna structural parameters on its performance, a simulation analysis is conducted, focusing on the influence of parameters S2, Kslot, and P on the return loss (S11) and realized gain. The simulation results are detailed in Fig. 5, specifically, Fig.5(a) to Fig.5(c) illustrate the effects of varying parameters S2, Kslot, and P on the return loss S11, respectively. Specifically, when P equals 2mm, its -10dB impedance bandwidth spans an extensive frequency range, reaching 24.4% (28.8-36.8GHz).

Fig. 5(d) specifically showcases the impact of parameter P on the realized gain. Through comprehensive analysis of the data presented in Fig. 5, it is evident that these parameters have a significant and sensitive influence on antenna performance. The final optimized parameters of the antenna (units: mm) are as follows: Lgx=11.6, Lgy=8, Lslot=6.4, Wdia=4.9, Gslot=2.8, Kslot=0.9, P=2, S1=0.1, S2=1.2, Ditop=0.3, Lport=0.6, H1=0.762, H2=1. Fig. 5(d) shows maximum realized gain of up to 8.4dBi. Despite its compact overall dimensions of 11.6mm x 8mm x 1.762mm, this antenna exhibits exceptional performance. The results underscore the antenna's remarkable ability to maintain both high-bandwidth and high-gain characteristics within a miniaturized form factor.

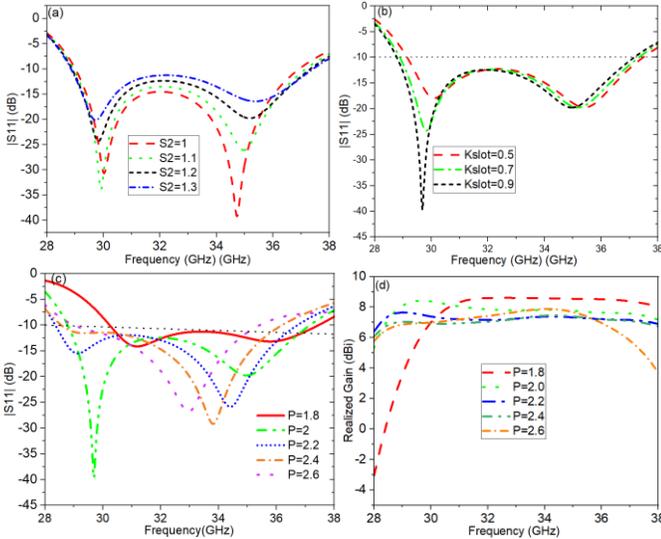

**Fig. 5.** the influence of parameters on the return loss (S11) and realized gain. (a) S2. (b) Kslot. (c) and (d) P.

### D. Antenna with PCMs

To reduce the RCS value of the antenna, the yellow elliptical PCM units and their mirror units are arranged in a checkerboard pattern on the top of the layer1, as shown in Fig. 6. This checkerboard structure is composed of 2x2 sub-periodic cells arranged in the different direction. This ultimately forms a checkerboard array with an overall size of 24mm x 40mm. After adding PCM around the antenna, coupling effects inevitably arise between the antenna and PCM, which may impact the antenna's performance. Therefore, the PCM units on the top area of SIW cavity are removed in Fig. 6 to mitigate the coupling effects between the antenna and PCM.

To facilitate the addition of SMA connectors for processing and testing, a feed line section has been incorporated into the antenna structure, as shown in Fig. 6. Fig.6(a) presents a 3D model. Side view is depicted in Fig. 6(b). Fig. 6(c) and Fig. 6(d) depict the configuration at the top and bottom of dielectric substrate layer1. The white through-holes are designed to securely fasten the two dielectric layers together using screws, ensuring that EM signals can pass through the entire dielectric board smoothly and without loss.

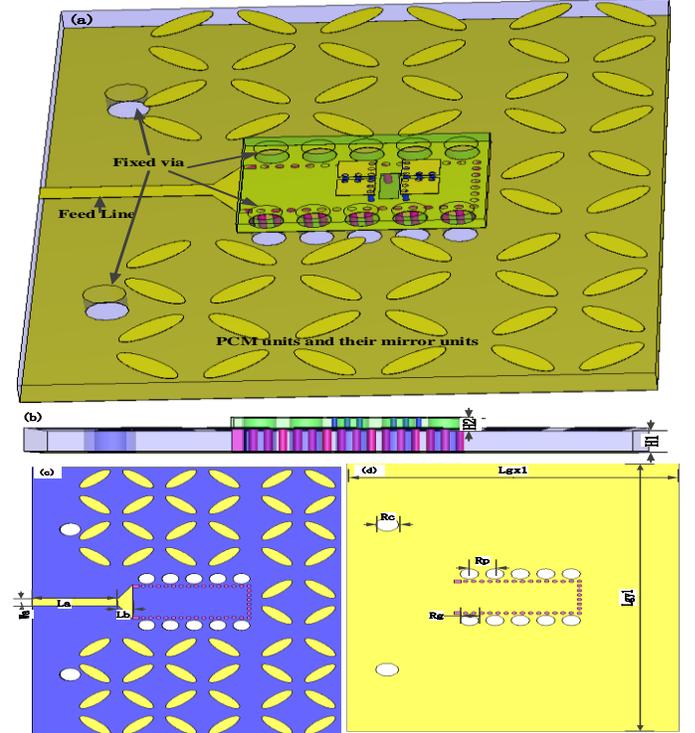

**Fig. 6.** Schematic diagram of antenna structure. (a)3D model antenna. (b)side view. (c)The top of dielectric substrate layer1. (d)The bottom of layer1.

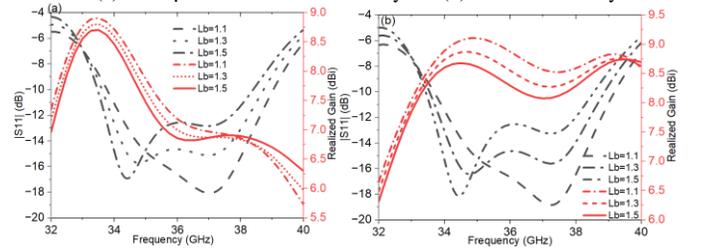

**Fig. 7.** The impact of parameter Lb on S11 and realized gain. (a) the reference antenna. (b) the proposed antenna

To delve into the specific impact of the impedance matching section length Lb on antenna performance, simulations are conducted for three scenarios with Lb set at 1.1mm, 1.3mm, and 1.5mm, with the results presented in Fig. 7. Fig. 7(a) displays the antenna without PCM (Ref. Ant.), while Fig. 7(b) shows the antenna with PCM (Pro. Ant.). Both antennas maintain consistent volume and area. In the Fig. 7, the black curves represent the S11, and the red curves represent the realized gain. By comparing the results of S11 and realized gain, it can be observed that when Lb=1.1mm, the antenna exhibits superior performance. The other final optimized parameters of the antenna (units: mm) are as follows: H1=1.575, H2=0.787,

Rp=2.2, Rg=2, Rc=2, Wa=1.2, La=8.2, Lb=1.1, Lgx1=28.7, Lgy1=41.6. From the perspective of impedance bandwidth, the -10 dB impedance bandwidth of the reference antenna is 14.29% (33.8-39 GHz). For the proposed antenna, after incorporating PCM, the -10 dB impedance bandwidth increases to 15.09% (33.7-39.2 GHz). This means that the introduction of PCM enhances the impedance bandwidth by 0.8%, although the improvement is relatively small, it still indicates that PCM has a positive effect on S11 performance.

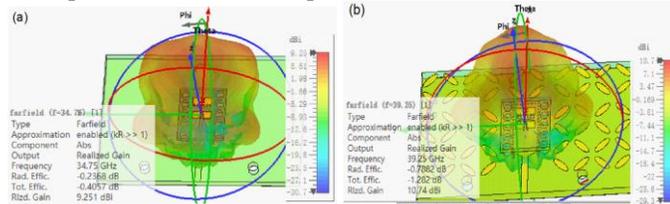

**Fig. 8.** Simulated 3D radiation patterns. (a) the reference antenna. (b) the proposed antenna

In terms of realized gain in the Fig. 7, the gain of the reference antenna decreases after 33.5 GHz, whereas the proposed antenna maintains a gain above 8.5 dBi when the frequency exceeds 33.5 GHz, with a maximum realized gain reaching 9.1 dBi. This further demonstrates that the proposed antenna has a more stable radiation gain. Fig.8 presents the 3D radiation patterns for two types of antennas. Specifically, the reference antenna achieves a peak realized gain of 9.25 dBi at 34.75 GHz, as shown in Fig. 8(a). In contrast, the proposed antenna reaches a peak realized gain of 10.74 dBi at 39.25 GHz, as illustrated in Fig. 8(b). This implies that after loading PCM, the realized gain of the antenna is improved by 1.49 dB.

## III. ANTENNA PERFORMANCES

### A. Radiation performance of proposed antenna with PCMs

To comprehensively evaluate the radiation characteristics and impedance bandwidth performance of the designed antenna, using standard PCB technology, a prototype of the antenna was ultimately crafted and subjected to practical measurements in an anechoic chamber, with the purpose of eliminating any potential interference from the surroundings and ensuring its alignment with theoretical predictions. The detailed picture of the antenna is depicted in Fig. 9, with dimensions of 28.7 mm × 41.6 mm × 2.362 mm. Fig. 9(a) showcases the fabricated prototype of the antenna. Fig. 9(b) comprehensively reveals the antenna structure of the layer2, including detailed designs of both its top and bottom. Fig. 9(c) presents a top view of the layer1 of the antenna. Fig. 9(d) displays the bottom configuration of the layer1. Fig. 9(e) illustrates the actual environmental layout for far-field measurements, offering context for understanding the measurement conditions.

To more intuitively compare the performance differences among various antenna structures, a comparative analysis of return loss and realized gain is conducted for the reference antenna, the designed antenna, and the physical antenna, with the comparison results presented in Fig. 10(a) and (b), respectively.

The simulations and measurements are in good agreement.

The measured and simulated −10 dB impedance bandwidths are 33.7-39.2 GHz and 33.76–39.84 GHz, respectively. To summarize, the operational bandwidths of the proposed antenna and the reference antenna are comparable. The measured relative bandwidth is approximately 16.52%. Similarly, for both measured and simulated data, the realized gain surpasses 8 dBi within the ranges of 33.24–38.52 GHz and 32.96–40 GHz, respectively. At 35.24 GHz, the peak measured realized gain reaches 8.9 dBi, whereas the simulated peak realized gain attains 9.12 dBi at 35 GHz, indicating a 0.22 dB discrepancy between measurement and simulation. Beyond 38 GHz, a decline in realized gain is observed, primarily attributed to factors such as assembly inaccuracies and measurement uncertainties.

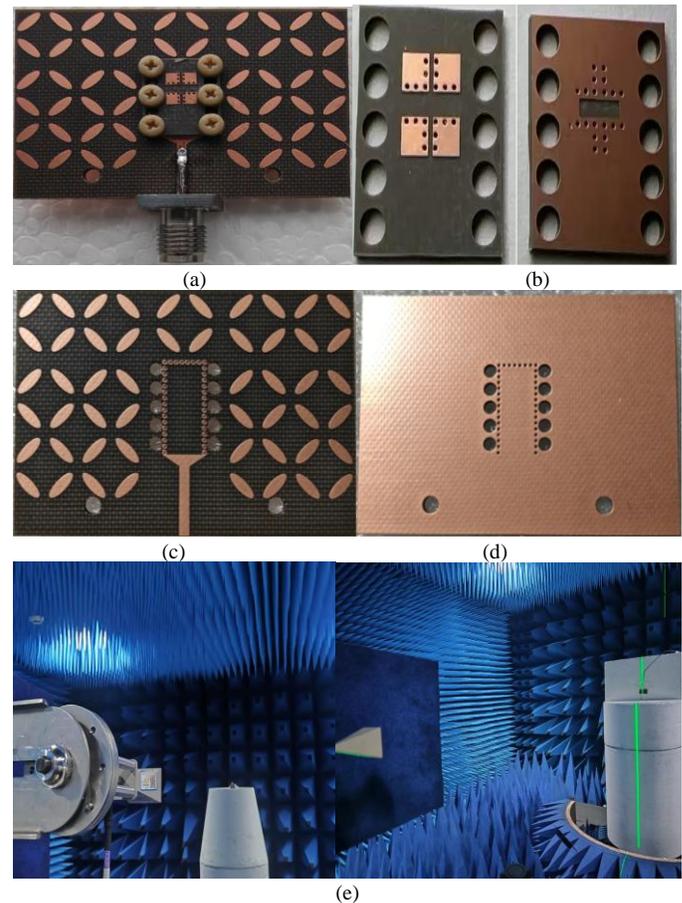

**Fig. 9.** Pictures of the proposed antenna. (a) Fabricated prototype. (b)The top of layer1. (c)The bottom of layer1. (d)The top and bottom of layer2. (e) Radiation performance measurement system.

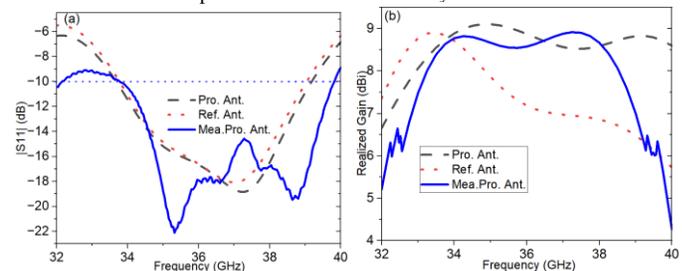

**Fig. 10.** Comparisons of the proposed antenna, reference antenna and physical model of the proposed antenna. (a) |S11|.(b) Realized gain.

## B. Scattering Performance of proposed antenna with PCMs

To comprehensively and accurately evaluate the scattering characteristics of the proposed antenna, a comparative result of RCS is conducted with a reference antenna of the same size and a metal plate. Fig. 11(a)-(c) display the three-dimensional bistatic RCS of these three types of antenna at 30 GHz (out-of-band) under normal incidence conditions; correspondingly, Fig. 11(d)-(f) showcase their RCS at 37.75 GHz (in-band).

Upon observing Fig. 11(a) and (d), it is evident that the metal plate exhibits strong scattering characteristics in the beam direction. In contrast, due to its inherent scattering effect, the reference antenna (as shown in Fig. 11(b) and (e)) scatters a portion of the incident wave's energy around the antenna, resulting in a reduced reflected signal intensity in the beam direction compared to the metal plate.

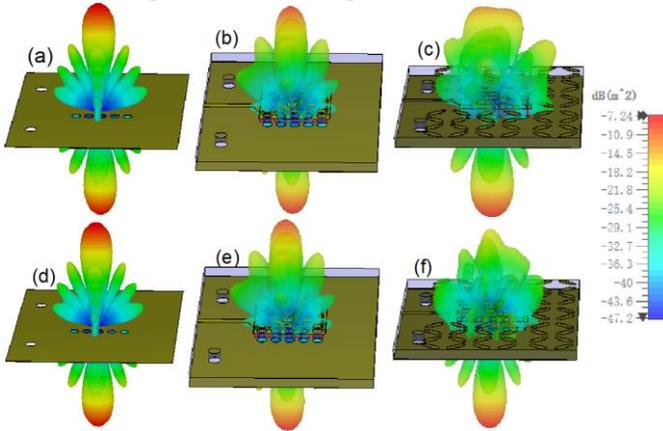

**Fig. 11.** Three-dimensional bistatic RCS under normal incidence. (a)metal plane, (b) the reference antenna and (c) the proposed antenna at 30GHz. (d) metal plane, (e) reference, and (f) proposed array at 37.75GHz.

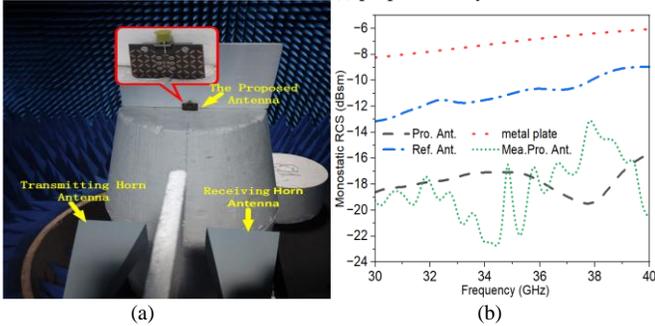

**Fig. 12.** Measurement and comparative chart of RCS. (a) Schematic view of measurement. (b)Comparative chart of monostatic RCS reduction effect of the proposed antenna, reference antenna and physical model of the proposed antenna.

When the frequency is within the band (37.75 GHz), the reference antenna not only exhibits scattering but also, owing to its good impedance matching characteristics, further enhances the energy scattered around it compared to the out-of-band scenario (as depicted in Fig. 11(b)), with a corresponding reduction in the reflected signal in the beam direction. As for the proposed antenna (shown in Fig. 11(c) and (f)), it not only possesses the scattering and impedance matching characteristics of the reference antenna but also, due to the integration of PCM, generates a phase cancellation effect between adjacent PCM units. This phase cancellation effectively redirects the incident wave into four different directions, significantly altering the antenna's radiation and scattering characteristics. Notably, this design substantially reduces the reflected energy in the beam direction of the antenna, successfully achieving RCS reduction.

Fig. 12 (a) presents a detailed schematic of the experimental setup for measuring the monostatic RCS reduction characteristics in a millimeter-wave anechoic chamber. This experiment employs two horn antennas, one serving as the transmitting antenna to emit plane waves that impinge on the designed antenna in a normal incidence manner, and the other acting as the receiving antenna to capture the reflected signals. This configuration covers a broad frequency range from 30 GHz to 40 GHz. Due to the phase cancellation properties of the PCM and the scattering characteristics of the antenna, most of the EM waves are scattered into the surrounding space or effectively absorbed upon encountering the target, with only a minimal portion of the signals being reflected back to the receiving antenna. Through meticulous analysis of the signals captured by the receiving horn antenna, an accurate RCS reduction value can be calculated, enabling the assessment of the stealth performance of the designed antenna. Fig. 12(b) presents a comparison of simulated and measured data for the monostatic RCS under normal incidence conditions. Within the frequency range of 30 to 40 GHz, the RCS reduction of the metal ground plane ranges from -7.77 to -6 dBsm, while the reference antenna exhibits an RCS reduction between -11.74 and -8.9 dBsm. The proposed antenna, however, achieves an RCS reduction of -19.67 to -15.6 dBsm. Compared to the metal ground, the proposed antenna demonstrates the maximum RCS reduction at 37.77 GHz, with a value of 13.076 dB, and the minimum reduction at 33.63 GHz, with a value of 9.74 dB. The impedance bandwidth of the antenna tested at -10dB is within the range of 33.76–39.84 GHz. However, as shown in Fig. 12, it achieves effective RCS reduction across the entire frequency band of 30-40 GHz. Furthermore, the PCM unit demonstrates a wide polarization conversion bandwidth of 25.3-59.3 GHz, indicating that the antenna not only achieves RCS reduction in band but also maintains good RCS reduction performance out of band. It is worth noting that there is a significant difference between the simulation results and the measured data, primarily due to inconsistencies between the simulation and actual measurement environments, as well as factors such as processing accuracy and manufacturing errors.

Table I presents a comparison with other low RCS antennas to more clearly demonstrate the advantages of the proposed antenna. Specifically, [10] focuses on the reduction of out-of-band RCS, while [11] and [12] concentrate on in-band RCS reduction. Furthermore, the unit sizes in [21] and [22] are relatively large, making it difficult to achieve miniaturization. In contrast, our design achieves a smaller unit size while maintaining performance, thus meeting the requirements for miniaturization.

## IV. CONCLUSION

This paper proposes a broadband antenna design that not only exhibits low RCS characteristics but also achieves high gain. The antenna ingeniously employs a checkerboard layout consisting of PCM units and their mirror units. Based on the principle of phase cancellation, it successfully achieves a significant RCS reduction of up to 13.076 dB within the Ka-band frequency range of 30-40 GHz. By integrating advanced RCS MS technology, the antenna maintains a low profile while significantly reducing its scattering properties. These unique advantages render it valuable for applications in the field of wireless communication and bode well for its broad development prospects.

TABLE I
COMPARISON WITH OTHER ANTENNAS

| Ref. | RCS Reduction Bandwidth | Element Size (units: mm) | RCS Reduction Method |
|---|---|---|---|
| [10] | out-of-band, 80.78%(4.96-5.6GHz),8 3.68%(6.88-7.43GHz). (>10dB) | 15.7 x15.7 x0.5 | AMCs and Reconfigurable technology |
| [11] | In-band, 29.2%8. (2-11GHz),>10 dB | 13.5 x13.5 x3.5 | Reconfigurable technology |
| [12] | In-band, 12.38%(9.85–11.15 GHz).(>10dB) | 15 x15 | Reconfigurable technology |
| [24] | In-band and out-of-band, 106%(6.10-18.9GHz) (>10dB) | 8.5 x8.5 x4.5 | PCM |
| [25] | In-band and out-of-band, 72.3%(7.6-16.2 GHz), (>10dB) | 13.5 x13.5 x3 | PCM |
| This work | In-band and out-of-band, 28.57%(30-40GHz), (>10 dB) | 4 x 4x1.575 | PCM |

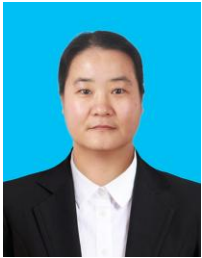

Cuiqin Zhao received B.S. degree in Electronic Information Science and Technology from Yunnan University, Kunming, China, in 2005, and M.S. degree in Communication and Information Systems from the same university in 2008. Afterwards, she worked as a full-time teacher at Hechi University in Guangxi, China. Her research interests include polarization conversion meta-surface, radar cross-section, frequency selective meta-surface, and antenna design.

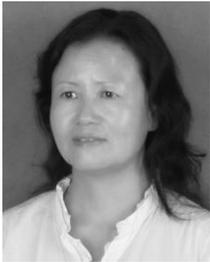

Dongya Shen (M12) received the B.S. degree from Yunnan University, Kunming, China, in 1985, and the M.Eng. degree in electrical engineering from the University of Electronic Science and Technology of China, Chengdu, China, in 1988. She is currently a professor of the communication and network technology with the School of Information Science and Engineering, Yunnan University, Kunming. Her current research interests consist of antennas of wireless communications, linearization of the radio over fiber systems, radio propagations, and channel modeling of wireless communications.

Ms. Shen was a recipient of the 2008 Third-Class Award from the Natural Science Prize of the Government of Yunnan Province, China. She is an Academic Leader of the Young and Middle-aged Talents of Yunnan Province, China, the Vice Chairman of the Microwave Integrated Circuits and the Mobile Communications Professional Committee of the Microwave Society of China, and a Committee Member of the Antenna Society of China.